\newcommand{\papertitle}{Dipole force free optical control and cooling
  of nanofiber trapped atoms}

\newcommand{\paperkeywords}{Quantum Optics, Cooling, Nanofiber,
  Nanooptics}

\makeatletter \def\namedlabel#1#2{\begingroup
  \def\@currentlabelname{#2}%
  \label{#1}\endgroup
} \makeatother

\documentclass[twocolumn,amsmath,amssymb, aps, showpacs, showkeys,
nobibnotes, floatfix, prl, 10pt, longbibliography]{revtex4-1}%

\usepackage[utf8]{inputenc} 
\usepackage[english]{babel} 

\usepackage{graphicx}
\usepackage{siunitx}%
\DeclareSIUnit\gauss{G} 
\usepackage{bm}
\usepackage{xspace}%
\usepackage{xpatch}
\usepackage{braket}%
\usepackage{wasysym}

\usepackage{xcolor}%
\definecolor{col0}{rgb}{0.000000,0.447000,0.741000}%
\definecolor{col1}{rgb}{0.850000,0.325000,0.098000}%
\definecolor{col2}{rgb}{0.929000,0.694000,0.125000}%
\definecolor{col3}{rgb}{0.494000,0.184000,0.556000}%
\definecolor{col4}{rgb}{0.466000,0.674000,0.188000}%
\definecolor{col5}{rgb}{0.301000,0.745000,0.933000}%
\definecolor{col6}{rgb}{0.635000,0.078000,0.184000}%

\newcommand{\linkcolor}{col0}%
\usepackage{hyperref} 
\hypersetup{colorlinks=true, %
  linkcolor=\linkcolor, %
  citecolor=\linkcolor, %
  filecolor=\linkcolor, %
  urlcolor=\linkcolor, %
  pdftitle=\papertitle, %
  pdfauthor={}, %
  pdfsubject={\paperkeywords}, %
  pdfkeywords={\paperkeywords} %
} %

\graphicspath{{images/}}

\makeatletter%
\def \@labelsection{%
  \@ifundefined{@sectioncntformat}%
  {\@seccntformat}%
  {\@sectioncntformat}{section}%
}%
\def \@labelsubsection{\@labelsection.\thesubsection}%
\def \@labelsubsubsection{\@labelsubsection.\thesubsubsection}%
\xpatchcmd{\@sect@ltx}{\@xsect}{%
  \let\@hskip\hskip%
  \def \hskip { \@hskip 0em plus}%
  \let\@MakeTextUppercase\MakeTextUppercase%
  \def \MakeTextUppercase{}%
  \edef \@currentlabelname{%
    \@hangfrom@section{}{\csname @label#1\endcsname}{#8}%
  } %
  \let\MakeTextUppercase\@MakeTextUppercase%
  \let\hskip\@hskip%
  \@xsect}{}{}
\makeatother

\newcommand{\wsb}{\ensuremath{\omega_\mathrm{sb}}\xspace}
\newcommand{\fref}[2][]{Fig.~%
  \ifx\\#1\\%
  \ref{#2}%
  \else%
  \ifdefined\hyperref{%
    \hyperref[#2]{\ref*{#2}#1}%
  }\else{%
    \ref{#2}#1%
  }\fi%
  \fi} 

\newcommand{\op}[1]{\ensuremath{\mathbf{\hat{ #1 }}}} %
\newcommand{\vect}[1]{\ensuremath{\mathbf{{ #1 }}}} %
\newcommand{\state}[1]{\ensuremath{{\vert #1\rangle}}} %
\newcommand{\trans}[2]{\ensuremath{\state{#1} \rightarrow}
  \state{#2}} %
\newcommand{\Dhfs}{\ensuremath{\Delta_\mathrm{hfs}}\xspace} %

\newcommand{\NBI}{QUANTOP, Niels Bohr Institute, University of
  Copenhagen, Blegdamsvej 17, 2100 Copenhagen, Denmark}

\begin{document}

\newcommand{\correspondingauthors} { %
  \email[Corresponding Authors: ]{jappel@nbi.dk}%
  \email{muller@nbi.dk}%
  \affiliation{\NBI}%
}

\title{\papertitle}

\author{C. {\O}stfeldt}%
\author{J.-B. B{\'e}guin}%
\author{F. T. Pedersen}%
\author{E. S. Polzik}%
\author{J. H. M{\"u}ller}\correspondingauthors%
\author{J. Appel} \correspondingauthors%

\date{\today}

\begin{abstract}
  The evanescent field surrounding nano-scale optical waveguides
  offers an efficient interface between light and mesoscopic ensembles
  of neutral atoms. %
  However, the thermal motion of trapped atoms, combined with the
  strong radial gradients of the guided light, leads to a
  time-modulated coupling between atoms and the light mode, thus
  giving rise to additional noise and motional dephasing of collective
  states. %
  Here, we present a dipole force free scheme for coupling of the
  radial motional states, utilizing the strong intensity gradient of
  the guided mode and demonstrate all-optical coupling of the cesium
  hyperfine ground states and motional sideband transitions. %
  We utilize this to prolong the trap lifetime of an atomic ensemble
  by Raman sideband cooling of the radial motion, which has not been
  demonstrated in nano-optical structures previously.  Our work points
  towards full and independent control of internal and external atomic
  degrees of freedom using guided light modes only.

  \ifdefined\svnid {%
    \begin{description}%
    \item[SVN] \footnotesize%
      \textcolor{red}{\svnFullRevision*{\svnrev} by
        \svnFullAuthor*{\svnauthor}, 
        Last changed date: \svndate }%
    \end{description}%
  } \fi%
\end{abstract}

\pacs{42.50.Ct, 37.10.De}

\keywords{\paperkeywords}
\maketitle

Light guided by nanooptical waveguide- and resonator structures
propagates partly as an evanescent wave; its tight sub-wavelength
confinement allows for strong interactions between guided light and
single atoms~\cite{Thompson2013,Goban2014,OShea2013,Kato2015} or
atomic ensembles~\cite{Vetsch2010,
  Goban2012a,Beguin2014,Lee2015,Corzo2016,Hunger2010,Haas2014} trapped
within the confined field.

The inherent intensity gradients of evanescent modes are a necessity
for the realization of dipole traps close to the surface of the
structure. However, if the atoms are probed or manipulated also by
evanescent modes, the gradients lead to detrimental effects, such as
time-dependent coupling for moving atoms, additional quantum partition
noise in probing atomic ensembles and motional dephasing of collective
internal quantum states. As strong gradients imply strong dipole
forces for any Stark shift induced by guided light, a scheme for
optical manipulation of the internal degrees of freedom without
perturbation of the motional state is desirable.

Additionally, any non-zero temperature above the motional quantum
ground state potentially decreases the average interaction of atoms
with the guided light mode, reducing the single atom optical depth
(OD).

Previous results for addressing these challenges in the nanofiber
platform \cite{Solano2017} include microwave cooling of the azimuthal
degree of freedom~\cite{Albrecht2016} by exploiting the
state-dependency of the trapping potentials for different Zeeman
sub-states, as well as polarization gradient
cooling~\cite{Goban2012a}.

\begin{figure}[tbp]
  \centering
  \includegraphics[]{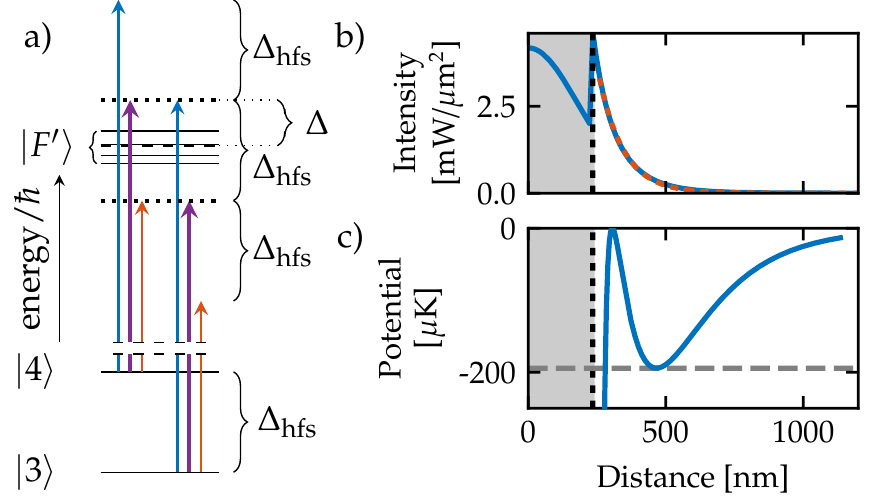}
  \caption{a) Stark shift canceling Raman scheme. The hyperfine ground
    states are coupled through two virtual levels (dotted lines)
    located symmetrically around the excited state manifold. The
    coupling fields are created by phase-modulation of a
    {\color{col3}carrier field}. By choosing the carrier detuning,
    $\Delta\approx\Dhfs/2$ and the ratio of the sideband to carrier
    powers $A= (J_1(z)/J_0(z))^2\approx 1.5$, the ac-Stark shifts are
    canceled.  A $\pi$ phaseshift of the {\color{col1}red-detuned
      sideband} compared to the {\color{col0}blue-detuned sideband}
    ensures constructive interference of the Raman coupling
    amplitudes.  b) Intensity of \SI{1}{\milli \watt} of light
    ($\lambda=\SI{852}{nm}$) guided by a nanofiber
    ($\diameter=\SI{470}{nm}$) as a function of distance to the fiber
    axis, together with an exponential fit (dotted line). c) Trapping
    potential for Cs atoms in the electronic ground state. See main
    text for details.}
  \label{fig:ramanScheme}
\end{figure}

In this paper, we present a Raman coupling scheme that allows us to
drive coherent transfers on the hyperfine transition in cesium (Cs) as
well as radial motional sideband transitions, while canceling all
quadratic ac-Stark shifts and thus dipole forces.  Driving Raman
transitions with a single beam propagating through the waveguide, we
implement a cooling protocol that relies on the gradient of the
coupling strength rather than its phase~\cite{kasevichchu}.

A key ingredient in our experimental implementation is the Stark shift
canceling Raman coupling scheme presented in
\fref[a]{fig:ramanScheme}: We couple the hyperfine ground states
\state{F=3} and \state{F=4} via two simultaneous two-photon Raman
processes. This is achieved by phase modulation (PM) of a single
optical field with a modulation frequency $\wsb=\Dhfs+\delta$, close
to the hyperfine ground state splitting~\Dhfs. The carrier field is
detuned $\Delta\sim\Dhfs/2$ above the transition from
\state{6^2S_{1/2},F=4} to the \state{6^2P_{3/2},\ F'} manifold.  By
choice of the absolute frequencies and powers of the fields, we cancel
the quadratic ac-Stark shifts.

To present the basic idea, we assume an initial field
$E(t)=E_0e^{i\omega t}$ subject to pure PM and neglect sidebands of
order two and higher
\begin{equation} %
  E(t) = E_0 J_0(z)e^{i\omega t}\bigg[1 + \sqrt{A}\Big( e^{+i\wsb t} -
  e^{-i\wsb t} \Big)  \bigg],
  \label{eq:mod}
\end{equation} %
where $A$~denotes the ratio of sideband-to-carrier power ${A =
  \left(J_1(z) / J_0(z)\right)^2}$, $z$~is the phase modulation index,
\wsb~is the modulation frequency, and $J_n$~is the $n^{\text{\tiny
    th}}$~ Bessel function of the first kind.

Neglecting hyperfine splitting of the upper state manifold, which is
much smaller than the single photon detuning $\Delta$, one obtains
simultaneous cancellation of the quadratic ac-Stark shifts for both
lower levels with $A=1.5$ and $\Delta=\Dhfs/2$. A more complete
numerical analysis of the parameter landscape including higher order
PM-sidebands and the excited state hyperfine splitting confirms that
$\Delta$ predominantly determines the common mode light shift, while
$A$ controls the differential light shift.

As the two (dominating) Raman couplings will add coherently, their
phase must be taken into account. Specifically, the transition through
the lower virtual level will acquire a minus-sign with respect to the
upper transition from the fact that the single-photon detuning is
opposite.  Constructive interference of the Raman coupling amplitudes
is ensured by the $\pi$-phase shift of the lower sideband relative to
the carrier and upper sideband inherent to PM -- see \eqref{eq:mod}.

The experimental setup is explained in detail
elsewhere~\cite{Beguin2014}, the salient features are summarized as
follows. A standard step-index fiber (Thorlabs 780HP) is tapered down
to sub-wavelength diameter ($d=\SI{470}{nm}$), so that light
propagating in the nanofiber is guided partly as an evanescent wave.
We create a dipole trap consisting of two quasi-linearly co-polarized
counter-propagating red-detuned beams ($\lambda=\SI{1056}{nm}$,
$P=2\times\SI{1.6}{mW}$) and one blue-detuned running wave field
($\lambda=\SI{780}{nm}$, $P=\SI{8.5}{mW}$) in the orthogonal
quasi-linear polarization~\cite{Vetsch2010}. The resulting trap
potential, evaluated at the axial minimum, is shown in
\fref[c]{fig:ramanScheme}. Axially, two strings of potential wells are
located on opposite sides of the fiber, allowing us to trap $\sim
2000$ atoms. Atoms are loaded into the trap by a standard 6-beam
magneto-optical trap (MOT) setup, followed by sub-Doppler
cooling~\cite{Vetsch2010,Beguin2014}.

We detect atoms in \state{F=4} using a dipole-force free dual-color
heterodyne dispersive measurement scheme previously described in
\cite{Beguin2014}. Atoms in \state{F=3} can be detected by transfer
into the \state{F=4}-manifold and subsequent measurement there. We
thus detect the optical phase shift induced by atoms in each state
onto our probe light and express the data in radians.

The Raman light is supplied from a standard extended cavity diode
laser, beatnote-locked to our MOT repump laser~\cite{Appel2009a}. This
leads to a flexible choice of carrier detuning, typically
\SI{4.5}{GHz} above the center of the transition from \state{F=4} to
the \state{F'=2\ldots 5} manifold.

The laser is modulated with by a {EOSpace PM-0K5-10-PFA-PFA-850}
fiber-coupled electro-optical modulator (EOM), driven by a RF signal
derived from a AD9910 DDS, which is mixed with a stable \SI{9}{GHz}
frequency~\cite{Louchet-Chauvet2010}. We measure the RF power at the
EOM, which is proportional to the squared modulation index~$z^2$.

The Raman light is coupled as a single (co-propagating) beam into the
fiber and is polarized parallel to the red trap light.

\begin{figure}[tbp]
  \centering
  \includegraphics[]{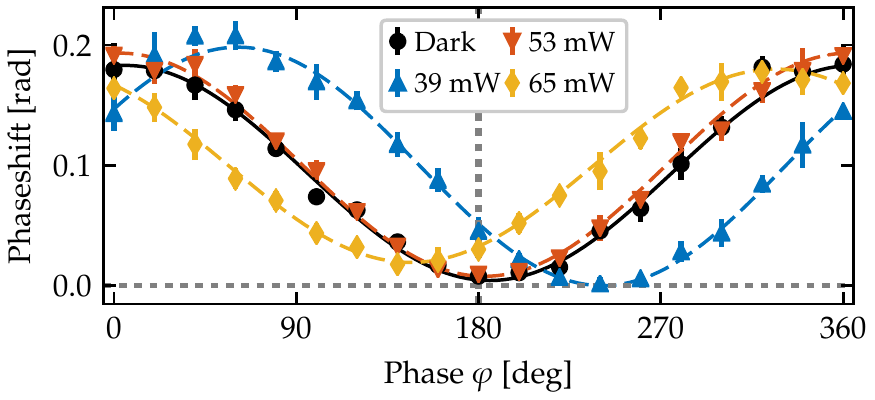}
  \caption{Ramsey fringes for varied sideband powers. An off-resonant
    Raman pulse perturbs the energy-splitting, leading to a shift of
    the fringe, $\varphi_0$. Black solid line: Ramsey fringe in the
    absence of the Raman pulse. Colored dashed lines: Fringes for
    varied EOM drive powers. For details see main text.  }
  \label{fig:starkshiftcancel}
\end{figure}

To set the phase modulation amplitude, we measure the differential
shift caused by the Raman laser in a Ramsey experiment in the
following way, see~\fref{fig:starkshiftcancel}: Starting with all
atoms in~\state{F=3,m_F=0}, the atoms are prepared in an equal
superposition $\left( \state{F=3,m_F=0} + \state{F=4,m_F=0}
\right)/\sqrt{2}$ by a microwave $\pi/2$-pulse. We then apply a
\SI{20}{\us} pulse of \SI{4.3}{nW} Raman light, where the PM phase is
flipped by \SI{180}{\degree} every \SI{4}{\nano \second} to avoid
population transfer. A second microwave $\pi/2$-pulse with
phase~$\varphi$ completes the Ramsey sequence, and the population in
\state{F=4,m_F=0} is measured. Any differential light shift induced by
the Raman light results in a shifted Ramsey fringe.  By adjusting the
sideband power ratios, we thus obtain cancellation of the differential
shift.

We verify that for the correct single photon detuning the Raman light
imposes no common mode ac-Stark shift, as we detect no modulation of
our probe signal for a Raman beam with approximately 1 order of
magnitude more power than used for normal coherent
operations~\cite{footnote}.

Coupling of motional states normally requires transfer of photon
momentum along the important direction, or alternatively
state-dependent potentials. For co-propagating Raman beams like in our
setup, the momentum transfer is only proportional to the wave-number
difference, and in our case it vanishes along the radial direction.
Instead, our scheme relies on the fast radial decay of the guided
Raman light, see \fref[b]{fig:ramanScheme}, for coupling the radial
motional states \state{n}:
\begin{align}
  \begin{split}
    \Omega_{n,n'} &= \Omega_0\Braket{ n | \frac{I(\op{r})}{I(0)} | n' }\\
    &\approx \Omega_0\Braket{n | \exp\left( - \op{r} / \ell \right) |
      n'},
  \end{split}
\end{align}
where $I(\op r)$ is the intensity at position $\op r$, $\ell$ is the
radial decay length of the intensity, $\Omega_0$ is the Raman Rabi
frequency for an atom located at the trap minimum $\op r=0$, and
\op{r} is the position-operator in the radial
direction~\cite{ChristofferThesis}. In analogy to the phase gradient
of a plane wave $\exp(i\vect k\vect r)$, the radial decay acts
effectively as ``imaginary momentum''. The coupling scales with the
ratio of the motional wavefunction size $\widetilde r$ to $\ell$.
This quantity plays the role of the Lamb-Dicke
parameter~\cite{Wineland1998}, and inserting for $\widetilde r$ the
ground state wave packet size we obtain $\widetilde r / \ell \approx
1/5$.

\begin{figure}[tbp]
  \centering
  \includegraphics[]{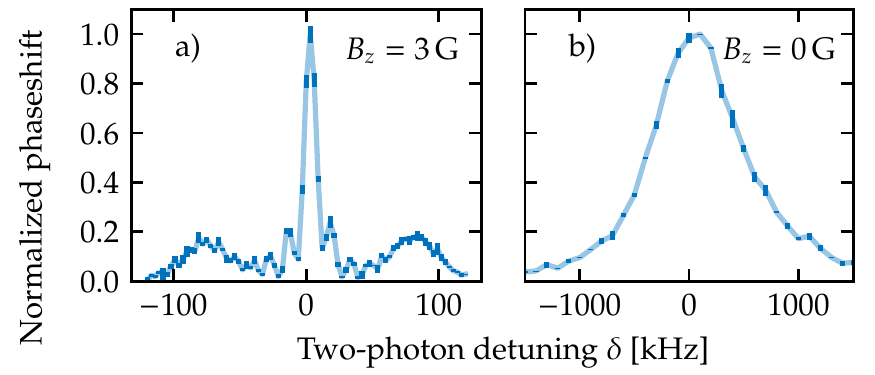}
  \caption{Typical Raman transfer spectra as a function of two-photon
    detuning.  a) Transfer spectrum with \SI{3}{G} bias field.
    \SI{80}{\us} pulse duration, leading to \SI{12.5}{kHz} oscillatory
    features seen close to the central peaks. b) Transfer spectrum
    without magnetic bias field. All values are normalized by the
    maximum value. Notice the different $x$-axes.}
  \label{fig:transferspectra}
\end{figure}

To investigate the Raman transfer efficiency, we prepare atoms in the
state \state{F=3, m_F=0}, and transfer them into \state{F=4,m_F=0}
using a resonant two-photon Raman pulse. In \fref{fig:transferspectra}
we show the transfer spectrum with and without a \SI{3}{G} magnetic
bias field, aligned parallel to the polarization of the blue-detuned
trap field.

In \fref[a]{fig:transferspectra} we observe for $\delta \equiv \wsb -
\Dhfs \sim \SI{0}{Hz}$ a motional carrier transition, i.e.~a
transition that changes only the internal state of the atom. For
$\delta\sim\pm(\SI{60}{kHz}-\SI{125}{kHz})$ we clearly resolve the
motional sidebands, demonstrating that we can couple the motional
states coherently. The sideband splitting is consistent with the
calculated frequency of the radial motion, and the width of the
sidebands stems from the spread of vibrational frequencies, due to the
anharmonic shape of the trap, see \fref[c]{fig:ramanScheme}.

In \fref[b]{fig:transferspectra} we plot a typical transfer spectrum
without the usual magnetic bias field. By removal of this field, the
different Zeeman states become degenerate, which we will utilize for
obtaining a simple repump scheme for cooling -- see below. Further, we
observe a significant broadening of the transfer spectrum without the
magnetic bias field. The line is broadened from essentially
interaction-time limited width, to $\sim\SI{1}{MHz}$ FWHM. At least
part of this broadening can be attributed to spatially varying vector
light shifts from the blue-detuned trap
light~\cite{Reitz2013,Lacroute2012}.  We note that the same broadening
is also observed in microwave transfer spectra under the same
conditions.

\begin{figure}[tbp]
  \centering
  \includegraphics[]{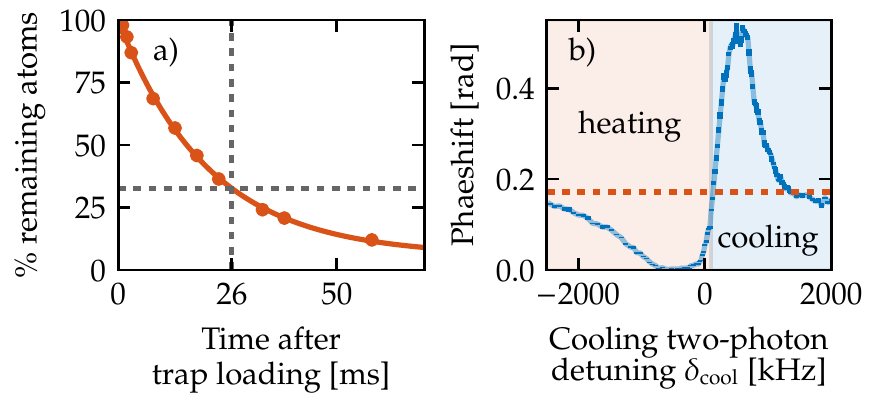}
  \caption{a) Remaining fraction of atoms in the trap without cooling
    as a function of time after trap loading. Solid line: Exponential
    fit.  Dashed lines indicate the atomic signal after a waiting time
    of \SI{26}{ms}. b) Atomic signal after 200 cooling cycles
    (duration \SI{26}{ms}), as a function of cooling two-photon
    detuning, $\delta_\mathrm{cool}$, (solid line, dark error bars)
    and atomic signal without cooling for equal time after trap
    loading (dashed line).}
  \label{fig:coolfreq}
\end{figure}

The ability to couple motional states opens the possibility to apply
Raman cooling of our ensemble, ultimately to the motional ground
state. Raman cooling to the ground state has been demonstrated in
nanoscale optical tweezer traps \cite{kaufmanprx,thompsonprl}.  In
nanofiber traps, a significantly increased background heating rate, as
compared to free space optical traps, is observed. This competes with
the Raman cooling process, and various mechanisms to explain this
effect have been put
forward~\cite{Vetsch2010,Wuttke2013,Lacroute2012}. Without cooling the
atomic population is heated from the trap, leading to a measured trap
lifetime of \SI{21}{ms}, see \fref[a]{fig:coolfreq}.  Similar values
have been reported for other nanofiber
traps~\cite{Goban2012a,Vetsch2010,Lee2015,Kato2015}.

We implement a Raman cooling scheme as follows: Our atoms are prepared
in the \state{F=4} level, without a magnetic bias field. A
\SI{40}{\us} Raman pulse transfers a fraction of the atoms to
\state{F=3}, with a concomitant decrease (increase) of the motional
quantum number for positive (negative) two-photon detuning of the
cooling light~$\delta_\mathrm{cool}$. After the Raman transfer, we
turn on the MOT repump laser resonant to the \trans{F=3}{F'=4}
transition for \SI{60}{\us}, which pumps atoms back into \state{F=4}
after on average 1.7 scattering events, significantly lower than the
number of scattering events needed for preparation of a pure Zeeman
state in the presence of a bias field. With 200 repetitions, the total
cooling sequence lasts \SI{26}{ms}, somewhat longer than the
$1/e$-lifetime of the atoms in the trap.  For cooling, the optical
power of the Raman beam and pulse duration were optimized for maximum
transfer efficiency on the sideband transition
\ensuremath{\state{F=3}\otimes\state{n} \rightarrow}
\ensuremath{\state{F=4}\otimes\state{n+1}} in the presence of a
magnetic bias field of \SI{3}{G}.

To assess the effect of cooling, we turn on the probe and MOT
repumper, and measure the remaining atoms, as shown in
\fref[b]{fig:coolfreq} as a function of $\delta_\mathrm{cool}$. For
$\delta_\mathrm{cool}\lesssim\SI{0}{kHz}$ we observe a clear reduction
of the atomic signal, whereas for
$\SI{100}{kHz}\lesssim\delta_\mathrm{cool}\lesssim\SI{1}{MHz}$ the
signal increases by up to a factor of $3.1$, compared to the case of
no cooling.

\begin{figure}[tbp]
  \centering
  \includegraphics[]{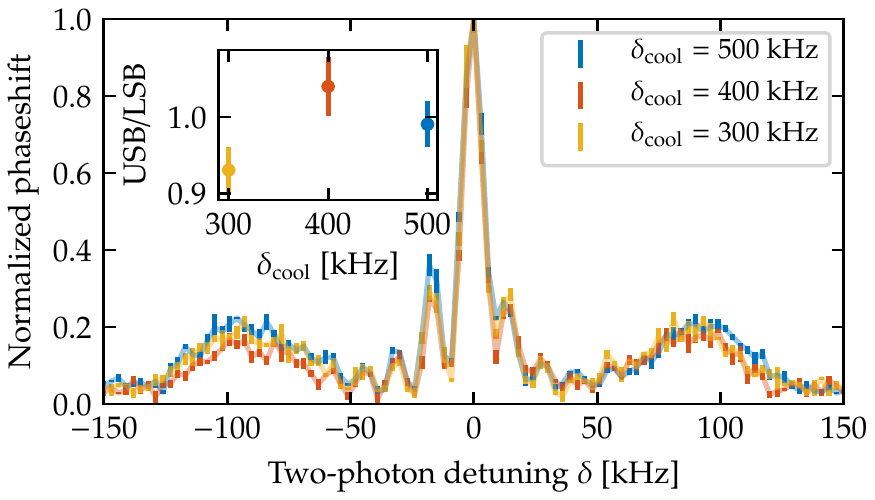}
  \caption{Raman transfer spectrum after 200 cooling pulses, for 3
    different values of two photon detuning during cooling,
    $\delta_\mathrm{cool}$. The spectra are normalized for clarity,
    due to an overall difference in atoms in the trap after the
    cooling sequence.  Inset: Ratio of the upper sideband (USB) to the
    lower sideband (LSB) integrated over the ranges
    $\pm(\SI{60}{kHz}$--$\SI{125}{kHz})$, as a function of
    $\delta_\mathrm{cool}$. Error bars show $1\sigma$ statistical
    uncertainties.} \label{fig:sidebandscooling}
\end{figure}

To further analyze the cooling performance, we perform the same
cooling sequence, after which we ramp up the bias field to \SI{3}{G}
and prepare the atoms in \state{F=3,m_F=0} by optical pumping and
microwave transfers~\cite{Tremblay1990}. We then record
sideband-resolved Raman spectra with an optical Raman laser power
similar to the one used in \fref[a]{fig:transferspectra}.  Example
spectra for three different values of $\delta_\mathrm{cool}$ are
displayed in \fref{fig:sidebandscooling}. The spectra show that we
cannot reliably detect a pronounced decrease in the radial temperature
of the atoms, heralded by a clear asymmetry between the upper and
lower motional sidebands. In the inset we show the integrated sideband
ratios (including $1\sigma$ statistical error bars). We observe a
slight indication of an increase in the ratio for
$\delta\sim\SI{400}{kHz}$, but the average motional quantum number
remains significantly above $\bar n = 1$.

Several effects are prone to reduce the efficiency of the cooling
scheme.  Working with a broadened Raman line at zero bias field
hinders the selective excitation of motional sidebands.  Secondly,
while the preparation of atoms into a specific state in the
\state{F=3}-manifold (e.g.~\state{F=3,m_F=0}) is necessary for the
resolved sidebands spectroscopy of the cooled ensemble, it comes at
the price of extra scattering events and extra time spent in the trap
after the cooling sequence, increasing the temperature. We anticipate
that in an improved setup a cooling protocol on a stretched level,
e.g.~\trans{F=3,m_F=3}{F=4,m_F=4} in the presence of a bias field can
be implemented. This scheme still allows for a simple and efficient
repump method, while motional sidebands are fully resolved.

We have presented a Raman coupling scheme for optical manipulation of
Cs atoms, that further cancels all quadratic ac-Stark shifts. We have
demonstrated coherent transfers using this scheme, as well as the
cancellation of the Stark shifts.

By utilizing the radial decay of the Raman laser light we have shown
that we can effectively couple the radial motional states by optical
manipulation.  We have further demonstrated first steps towards
experimental implementation of cooling of the radial degree of
freedom. We show that we can extend the lifetime of atoms in the trap
by application of 200 pulses of Raman cooling.

In summary, we have detailed an experimentally feasible way for
obtaining optical manipulation of neutral atoms around nanofibers
using exclusively guided light modes. Optical manipulation of atoms
trapped in nanoscale optical systems offers exciting new possibilities
such as position-dependent manipulation and adiabatic pulses on
timescales faster than the motional frequencies.

\begin{acknowledgments}
  Funding: ERC grant INTERFACE (grant no.~ERC-2011-ADG~20110209).
  The authors would like to thank Signe~B.~Markussen for help with data
acquisition.
\end{acknowledgments}

\bigskip

\bibliography{articlebib}

\begin{thebibliography}{25}
\bibitem [{foo()}]{footnote}%
  \BibitemOpen
  \href@noop {} {}\bibinfo {note} {The absolute cancellation point for atoms in
  \state{F=4,m_F=0} and \state{F=4,m_F=\pm4} is almost identical, as verified
  by explicit numerical calculations.}\BibitemShut {Stop}%
\end{thebibliography}

\end{document}